\documentstyle[amssymb,multicol,epsfig,aps,prl]{revtex}

\ifpreprintsty\def\multb{ }\def\multe{ } \else\def\multb{ \begin{multicols}{2}}\def\multe{ \end{multicols}} \fi

\begin{document}
\draft
\title{Robust half-metallicity and metamagnetism
in Fe$_{x}$Co$_{1-x}$S$_{2}$}
\author{I.I. Mazin}
\address{Code 6391, Naval Research Laboratory, Washington, DC 20375}
\date{January 19, 2000}
\maketitle

\begin{abstract}
The Fe$_{x}$Co$_{1-x}$S$_{2}$ system is predicted, on the basis of density
functional calculations, to be a half metal for a large range of
concentrations. Unlike most known half metals, the half metallicity in this
system should be very stable with respect to crystallographic disorder and
other types of defects. The endmember of the series, CoS$_{2},$ is not a half
metal, but exhibits interesting and unusual magnetic properties which can,
however, be reasonably well understood within the density functional theory,
particularly with the help of the extended Stoner model. Calculations
suggest strong electron-phonon and electron-magnon coupling in the system,
and probably a bad metal behavior at high temperatures.
\end{abstract}

\multb
\begin{flushright}
{\it God made the integers, all else is the work of man \\
(L. Kronecker). }
\end{flushright}
\vskip -.1in
Half metals (HM), materials that are metals in one spin channel and
insulators in the other, are attracting substantial interest recently, mostly
because of potential application in spintronics devices,
$e.g.$, spin valves\cite
{spintronics}. Unfortunately, although a few dozen various materials have
been predicted to be HM on the basis of band structure calculations, there
are hardly any that have been convincingly confirmed
to be such by an experiment. A possible
exception is CrO$_{2}$ where a spin polarization of up to 90\% was measured by
the Andreev reflection technique\cite{soulen} (in some other materials there
is indirect evidence such as integer magnetic moment or optical spectra
consistent with half-metallic bands). 
The usual explanation of such a discrepancy between the theory and
 the experiment is that half-metallicity in
these materials is very sensitive to crystallographic disorder and
stoichiometry (see, e.g., Ref. \cite{orgassy}). Indeed, no materials have
been predicted to be HM in a wide range of concentrations of the
constituents, and insensitive to disorder.

In this Letter I point out
one such material. Namely, I show that the
pyrite alloys Fe$_{1-x}$Co$_{x}$S$_{2}$ are HM for the most of the
concentration range (0.1$\lesssim x\lesssim 0.9$), and not sensitive to
ordering of Fe and Co in the metal sublattice. On the other hand,
at $x\agt 1$\cite{x1}, calculations predict magnetic collapse
under pressure, and a metamagnetic behavior just before the collapse.
Both half metallicity and metamagnetism can be explained by competition 
between the kinetic (band) energy, and the Stoner (Hund) interaction.

Experimentally, the system of pyrite solid solutions, (Fe,Co,Ni,Cu)(S,Se)$%
_{2},$ is amazingly rich. FeS$_{2}$ is a nonmagnetic semiconductor, in
agreement with the band structure calculations\cite{eyert,helmut}. With
as little as  0.1\% Co, by some data\cite{jarret}, it becomes a ferromagnetic
metal, and remains ferromagnetic
 all the way through CoS$_{2},$ and further on until approximately Co$%
_{0.9}$Ni$_{0.1}$S$_{2}.$ It was noted that the magnetic moment, $M,$ per Co
atom in Fe$_{1-x}$Co$_{x}$S$_{2}$ solid solutions stays close to 1 $\mu _{B}$
in an extremely wide range from $x\approx 0.1-0.2$
to $x=0.9-0.95.$ To the best of my knowledge, no
explanation of this fact has been suggested till now.
 At larger $x,$ $M$ decreases to $\approx 0.85$ $\mu _{B}$\cite
{jarret}. Starting from CoS$_{2},$ one can also
substitute S with Se.  The Curie temperature, $T_{C}$,
rapidly decreases, and magnetism disappears at
Se concentration of 10-12\%, while $M$ decreases only slightly\cite{se}. At
larger Se  concentrations, the material shows metamagnetic behavior with no
sizeable spontaneous magnetization, but with magnetization of 0.82$-0.85$ $%
\mu _{B}$ appearing abruptly when applied magnetic field in exceeds
approximately $(220x_{Se}-25)$ tesla.
A very similar behavior was observed in pure
CoS$_{2}$ under pressure\cite{se}, suggesting that the magnetic effect of Se
is just the density of states (DOS) reduction.

As shown below, 
all these effects are reproduced by the standard
local spin density (LSD) calculations, and find their explanation within the
extended Stoner model (ESM)\cite{eStoner}. 
Further substitution of Co by Ni leads to a Mott-Hubbard transition
into an antiferromagnetic state, which is, in
contrast to the system considered
here, poorly described by the LSD. Further doping by Cu leads to a
superconductivity, presumably due to high-energy sulfur vibrons\cite{ep}.

\begin{figure}[tbp]
\centerline{\epsfig{file=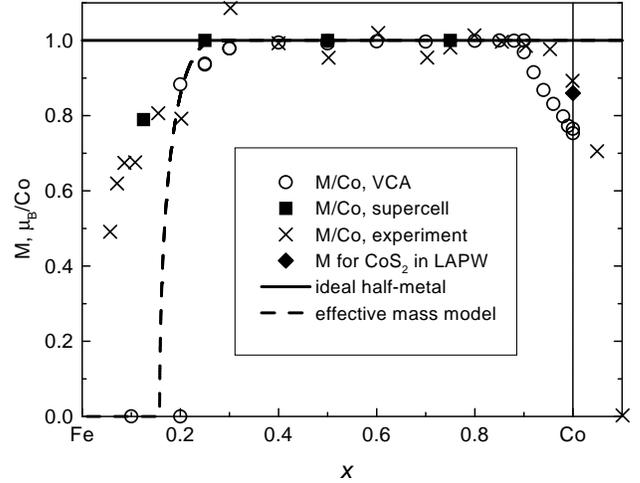,height=0.85\linewidth}}
\vspace{-.1in} \setlength{\columnwidth}{3.2in} \nopagebreak
\caption{ Experimental and calculated magnetic moment per Co
in Fe$_{1-x}$Co$_{x}$S$_{2}$ alloys. VCA: virtual crystal approximation,
s/cell: supercell calculations, experimental data are from
 Ref.\protect\cite{jarret}.
} \label{M(x)}
\end{figure}

To understand  the behavior of the Fe$_{1-x}$Co$_{x}$S$_{2}$ system, I
performed several series of density functional LSD calculations: First,
virtual crystal approximation (VCA) was used in conjunction with the Linear
Muffin Tin Orbital (LMTO) method\cite{lmto}. Then, I did several
calculations using rhombohedral supercells of 4 or 8 formula units. Finally,
I checked the results against more accurate
full-potential linear augmented plane wave
calculations\cite{wien} for pure CoS$_{2},$ FeCo$_{3}$S$_{8},$ and (Fe$%
_{0.25}$Co$_{0.75})$S$_{2}$ in VCA. The results were 
consistent with the LMTO calculations. All calculations were
performed in the experimental crystal structure of CoS$_{2}.$ In
reality, the S-S bond in FeS$_{2}$ is 4\%
longer. The effect on the band structure is not negligible, particularly
near the bottom of the conductivity band\cite{tbp}. However,
 the difference
is not important for the purpose of the current Letter, namely the half
metallicity of Fe$_{1-x}$Co$_{x}$S$_{2}$ alloys, and understanding the basic
physics of its magnetic phase diagram.
The resulting
magnetic moments are shown in Fig.\ref{M(x)}. In good
agreement with the experiment, the magnetic moment per Co is exactly 1 $\mu
_{B}$ for the Co concentrations 0.3$\lesssim x\lesssim 0.9.$ The same holds
for the rhombohedral supercell calculations (Fig.\ref{M(x)}%
), demonstrating stability of the HM state with respect to crystallographic
disorder. I performed calculations for ordered Fe$_{7}$CoS$_{16},$ Fe$_{3}$%
CoS$_{8},$ FeCoS$_{4},$ and FeCo$_{3}$S$_{8},$ and found that already Fe$_{7}$CoS$_{16}$
$(x=0.125)$
has magnetic moment of 1 $\mu _{B}$ per Co. Although the
original paper\cite{jarret} implied that the total magnetic
moment resides on Co,
this is not true: for instance, in Fe$_{7}$CoS$_{16}$ less than 30\% of the
total magnetization (0.45 $\mu _{B})$ resides on Co. The nearest neighbor Fe
(6 per cell) carry $\approx 0.15$ $\mu _{B}$ each. 8\% of the total moment
resides on S, about the same relative amount as in CoS$_{2}$\cite{S}. So,
one cannot view the low-doping Fe$_{1-x}$Co$_{x}$S$_{2}$ alloys as magnetic
Co ions embedded in polarizable FeS$_{2}$ background, as for instance Fe in
Pd.

The best way to understand the physics of this alloy is to start with FeS$_{2}.$ 
FeS$_{2}$ is a
nonmagnetic semiconductor with a gap between the $t_{2g}$ and $e_{g}$
states \cite{eyert,helmut}.
The reason for that is that sulfur forms S$_{2}$
dimers with the $pp\sigma $ states split into a bonding and an
antibonding state. The latter is slightly above the Fe $e_{g}$ states and
thus empty. The other 5 S $p$ states are below the Fe $t_{2g}$ states, hence
the occupation of the Fe $d$ bands is 6, just enough to fill the narrow $%
t_{2g}$ band. Magnetizing FeS$_{2}$ would require transfer of electrons from
the $t_{2g}$ into the $e_{g}$ band, at an energy cost of the band gap $%
\Delta $ per electron. The Stoner parameter $I,$ which characterizes the
gain of Hund energy per one electron transferred from the spin-minority into
the spin-majority band, appears to be smaller than $\Delta \approx 0.75$ eV
(For the conductivity band in FeS$_{2},$ due to hybridization with S, $I$ is
smaller that in pure Fe\cite{I} and is $\approx 0.55$ eV). However, if we
populate the same band structure with $x\ll 1$ additional electrons per
formula unit, we can either
distribute them equally between the two spin subband,
or place $x^{\prime }$ in the spin majority band and $x^{\prime \prime
}=x-x^{\prime }$ in the spin majority band (the total magnetic moment in $%
\mu _{B}$ is then $M=x^{\prime }-x^{\prime \prime }$).
In the latter case we gain a Hund energy of $-IM^{2}/4,$ but we 
lose kinetic (band structure)
energy because some states have only single occupancy and thus
more high-energy states have to be occupied. Using effective mass
approximation for the  conductivity band one can express this kinetic
energy loss in terms of concentration $x$ and  magnetic moment per Co
$\beta =M/x$ as $A(x/2)^{5/3}[(1+\beta )^{5/3}+(1-\beta
)^{5/3}]$, where $A=\frac{3\hbar ^{2}}{10m^{*}}(\frac{3}{4\pi V})^{2/3},$ $%
m^{*}$ is the effective mass, and $V$ is the volume
per formula unit. 

Minimizing the total energy with respect to $\beta $ gives the equilibrium
magnetic moment in ESM: 
$$
\beta  =(5A/3I)[(1+\beta )^{2/3}-(1-\beta )^{2/3}]/(4x)^{1/3}
$$
The solution is a universal function $\beta (Ix^{1/3}/A).$ Note that $\beta
(z)=0$ for $z<10\cdot 2^{1/3}/9$,
and $\beta (z)=1$ for $z>5/3.$ Correspondingly,
with doping the material remains paramagnetic until $x$ reaches $%
x_{1}=2(10A/9I)^{3},$ and then the magnetic moment per Co atom
gradually grows until the concentration reaches the
$x_{2}=(5A/3I)^{3}.$ At larger dopings the material remains
half-metallic with $\beta =1.$ Eventually, DOS starts to
deviate from the effective mass model and the polarization $\beta $ may
become smaller than 1 again. One can estimate the critical concentrations $%
x_{1,2}$ using the following values, extracted from the band structure
calculations for FeS$_2$: $m^{*}\approx 0.8,$ $I\approx 0.55$ eV. This gives $%
x_{1}\approx 0.16$ and $x_{2}\approx 0.26,$ in qualitative agreement with
the experiment and LSDA calculations (Fig.\ref{M(x)}).

This explains the on the first glance unexpected result: in contrast to most
known half metals Fe$_{1-x}$Co$_{x}$S$_{2}$ remains a HM for a large range
of concentrations, is insensitive to crystallographic disorder, and probably
not very sensitive to the state of the surface either: the behavior
qualitatively described by the universal function $\beta $ above is
determined by the competition of two large energies: The band gap $\Delta $,
which is 
a measure of crystal field splitting, and the Stoner factor $I$, a
measure of the Hund coupling. As long as $\Delta >I$ and $x_{1}\ll 1$,
a large region of half-metallicity exists. Both conditions are related
primarily to the atomic characteristic of constituents and gross features of
the crystal structure, and are not sensitive to details.

The same competition between the band energy and the Stoner energy
leads to very different magnetic properties
in case of stoichiometric CoS$_{2}
$, and of CoS$_{2}$ doped with Ni ($x>1$ region in Fig.1). The
experimental moment in the stoichiometric compound is 0.85-0.9 $\mu
_{B}$, in perfect agreement with full potential LSDA calculations, and in
reasonable agreement with LMTO results as well. External pressure, or
substituting S by Se, rapidly reduces magnetic moment\cite{se}. As shown in
Ref.\cite{Sebands}, Se doping increases the width of the conductivity
band (because the Se$_{2}$ dimer has smaller $pp\sigma $
splitting than S$_{2}$, and therefore the Co $e_{g}$ states are better aligned
with the chalcogen $pp\sigma ^{*}$ states\cite{tbp}). Thus Se doping has the
same effect as applying pressure. Doping with Ni has similar effect and
Ni concentrations of the order of 10\% make the system non-magnetic.
The large magnetic moment of CoS$_{2}$ suggests that the ferromagnetism in this
system is robust and the fact that it disappears so rapidly with pressure
and doping seems surprising.
\begin{figure}[tbp]
\centerline{\epsfig{file=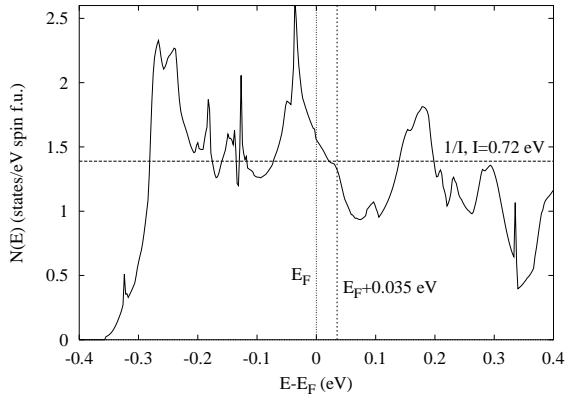,width=0.9\linewidth}}
\vspace{0.1in} \setlength{\columnwidth}{3.2in} \nopagebreak
\caption{Density of states for {\it paramagnetic} CoS$_2$, calculated
by LAPW method. $E-E_F=0.35$ eV corresponds to Co$_{0.9}$Ni$_{0.1}$S$_2$
in the rigid band approximation.
} \label{CoS2-DOS}
\end{figure}
Since all these effects are reproduced very well
by regular LSDA calculations (Fig.1), ESM
calculations again provide valuable physical insight.
Unlike the parabolic band case considered above, now one has to
take into account specific structure of DOS of CoS$_2$ (Fig.\ref{CoS2-DOS}).
The loss of the one-electron energy can be
expressed\cite{N(M)} in terms of the
 average DOS, $\tilde{N}(M),$ defined as $
M/(\mu _{B}H_{xc})$, where $(\mu _{B}H_{xc})$ is the (rigid) exchange splitting
producing magnetic moment $M$:
$\Delta E=\int_{0}^{M}MdM /2\tilde{N}(M)$.
 The best visualization of the ESM is via plotting $%
\tilde{N}(M)$ as a function of $M$. Wherever this curve crosses the line $1/I
$, one has an extremum of the total energy. If the slope of the curve is
positive at the intersection point, the extremum is a maximum, and the state is
unstable, otherwise it
is a minimum and indicates a (meta)stable magnetic state. The ESM plot for
CoS$_{2},$ based on the LAPW DOS (Fig.\ref{CoS2-DOS}) is shown on Fig.\ref
{esm}. At normal pressure, there are potentially two metastable states: a
low spin, of the order of 0.3 $\mu _{B}$, and a high spin, of the order of
0.85 $\mu _{B}$. For Stoner parameter $I<0.63$ eV neither state is
(meta)stable, for $0.63<I<0.71$ eV only the low-spin state is, for $%
0.70<I<0.73$ eV there are two metastable states, and for yet larger $I$ only
the high spin state may be realized. For $I>0.77$ eV ESM produces a HM. One can
estimate $I$ for a compound using the procedure described in Ref.\cite{MS},
or deduce it by comparing the ESM to fixed moment calculations. The latter
method leads to $I=0.68$ eV for LMTO and $I=0.72$ eV for full potential
calculations. The complicated structure of $\tilde{N}(M)$ can be traced down
to the structure of DOS near the Fermi level (Fig.\ref{CoS2-DOS}):
 after the initial drop of $\tilde{N%
}(M),$ it starts to increase again when the initially fully occupied 
peak at $E-E_{F}\approx -0.1$ eV becomes magnetically 
polarized. The high-spin solution
corresponds to the situation where this peak is fully polarized.

\begin{figure}[tbp]
\centerline{\epsfig{file=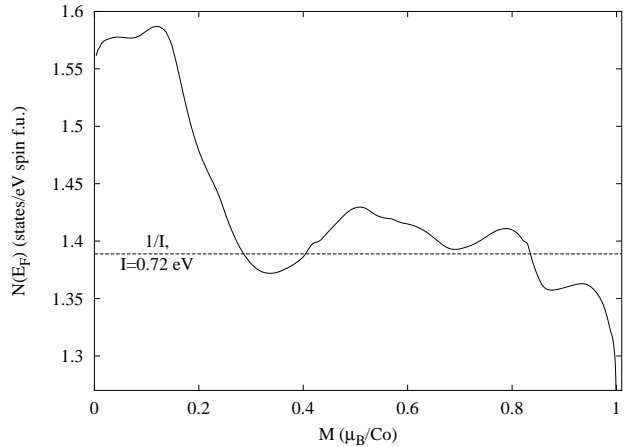,height=0.7\linewidth}}
\vspace{0.1in} \setlength{\columnwidth}{3.2in} \nopagebreak
\caption{Extended Stoner plot for CoS$_2$, showing the effective 
(averaged) density of states $\tilde N(E_F)$ as a function of
magnetic moment $M$. The horizontal line corresponds to a Stoner factor $
I=0.72$ eV. The two crossing points near $M\approx 0.3$ $\mu_B$, and
$M\approx 0.8$ $\mu_B$, show two (meta)stable states in
the rigid band approximation.
} \label{esm}
\end{figure}
Let me now consider the effect of pressure.
Roughly speaking, applying pressure amounts to rescaling the
band structure proportionally to squared inverse lattice parameter.
Correspondingly, the density of states is rescaled proportional to $%
(V_{0}/V)^{2/3}.$ The total energy in the ESM will include that as 
\begin{eqnarray}
E(V,M)&\approx& E_{0}+\left( \frac{V}{V_{0}}\right) ^{2/3}\int_{0}^{M}\frac{MdM%
}{2\tilde{N}(M,V_{0})} \nonumber\\
&-&\frac{IM^{2}}{4}+\frac{B(V-V_{0})^{2}}{2V_{0}},
\label{EOS}
\end{eqnarray}
where $V_{0},$ $E_{0},$ and $B$ are the equilibrium volume, energy, and the
bulk modulus, respectively, of the paramagnetic phase. The magnetic energy
shifts the energy minimum towards larger volumes, a standard
magnetostriction effect. What is unusual about the equation of states (\ref
{EOS}) with $\tilde{N}(M,V_{0})$ from Fig.\ref{esm} is that for some range
of values of $I$
 there
are two local minima, a high spin state with a larger volume, and a low spin
state with a smaller volume. 
The actual $I$ seems to fall into this range.
This suggests a first order phase transition
with pressure, and there are indications\cite{1O}
 that it has been observed in the
experiment\cite{se}, although the low spin state, which has  $M\approx
0.2$ $\mu _{B}$ in ESM and $\approx
0.1$ $\mu _{B}$ in fixed moment calculations,
was reported in Ref. \cite{se}
to have no or very small magnetic moment. 
Another consequence of the physical
picture outlined here is metamagnetism: In the low spin state close to the
critical pressure the system can be switched over to the high-spin state by
an external magnetic field defined by the energy  and magnetic
moment difference between the
two states. Again, metamagnetic
behavior has been observed in Co(S,Se)$_{2},$ in Co$_{1-x}$Ni$_{x}$S$_{2},$
and in compressed CoS$_{2}.$ Finally, the theory predicts rapid increase of
the equilibrium magnetization with {\it negative} pressure: in the ESM an
expansion of 3-4\% in volume already increases $M$ to nearly 1 $\mu _{B}.$
This suggests that the spin fluctuations at high temperature (above $T_{C}$)
may have larger amplitude than the ordered moments at zero temperature. This
kind of behavior has also been observed\cite{M(T)}, and discussed in the
literature\cite{Ogawa83}.

The family of pyrite materials formed by 3d transition metals and chalcogens
is incredibly rich. It shows various kinds of magnetism and metamagnetism,
metal-insulator transitions of different types, superconductivity, and half
metallicity. Except for the vicinity of a Mott-Hubbard transition, that is,
close to NiS$_{2},$ the physics of these materials can be rather well
understood within the local spin density functional theory. In particular,
the extended Stoner formalism provides considerable insight into the magnetic
behavior of this system. The most important conclusions from the
calculations are:

1. The Fe$_{1-x}$Co$_{x}$S$_{2}$ alloy is predicted to be a half metal in a
large range of concentration. Unlike most other known half metals, and very
importantly for the applications, the half-metallicity should be robust with
respect to defects and crystallographic disorder.

2.  CoS$_{2}$ is an itinerant ferromagnet which has, due to a
complicated structure of its density of states, two magnetic states: a 
high-spin one, with the moment $M\sim 0.8$ $\mu _{B},$ and a low-spin one,
with $M\alt0.1$ $\mu _{B}.$ The first order transition to the low spin state
can be induced by external pressure, doping with Se, or with Ni. In the low
spin state the material exhibits metamagnetic properties. The structure of
the density of states also manifests itself via an unusual temperature
dependence of the magnitude of magnetic fluctuations.

3.  The closeness of CoS$_{2}$ to a magnetic phase
transition also leads to strong coupling between electronic, lattice, and
magnetic degrees of freedom, making it a relative of such ``bad metals'' as
SrRuO$_{3}$ and magnetoresistive manganites. Therefore interesting transport
properties are to be expected in the system at and near stoichiometric
composition, including large magnetoresistance and violation of the
Yoffe-Regel limit at high temperatures.

For such an interesting, the 3d pyrites seem to be unusually
little studied either experimentally or theoretically. Hopefully this paper
will encourage further investigations.

\multe

\begin{references}
\bibitem{I}  Regarding calculations of the Stoner factor $I$ for compounds,
the reader is referred, for instance, to Ref.\cite{MS}
\bibitem{N(M)}This formula follows from the fact the 
derivative of the one-electron energy with respect to $M$ is 
$(\mu _{B}H_{xc})/2=(M/2)/\tilde{N} (M)$.
\end{references}
\end{document}